\documentclass[twocolumn,prl,showpacs,showkeys,amsmath,amssymb]{revtex4}
\usepackage{graphicx}
\usepackage{dcolumn}
\usepackage{bm}
\usepackage{ulem}
\begin{document}
\title{Phasor analysis of atom diffraction from a rotated material grating}
\author{Alexander D.\ Cronin, John D. Perreault}
\affiliation{University of Arizona, Tucson, Arizona 05721}
\date{\today}

\begin{abstract}
The strength of an atom-surface interaction is determined by
studying atom diffraction from a rotated material grating. A
phasor diagram is developed to interpret why diffraction orders
are never completely suppressed when a complex transmission
function due to the van der Waals interaction is present. We also
show that atom-surface interactions can produce asymmetric
diffraction patterns. Our conceptual discussion is supported by
experimental observations with a sodium atom beam.
\end{abstract}
 \maketitle

Atom diffraction from a material grating has recently been used to
measure the strength of van der Waals interactions between atoms
and the grating.  This was possible in \cite{gris99,gris00,bruh02}
because atom-surface interactions modify the intensity in each
diffraction order as a function of atomic $velocity$.  In this
paper we measure van der Waals interactions by studying
diffraction as a function of incidence $angle$. We also
demonstrate an asymmetric atom diffraction pattern from a
fabricated material grating.

Historically, a graphical analysis in the complex plane has been
useful to understand optical diffraction.  This is especially true
for the Fresnel integrals for which no closed form analytical
solutions have been found, yet the Cornu spiral permits a physical
interpretation \cite{Hecht}.  We have adapted this approach to our
current problem in atom optics.  Even the far-field limit, van der
Waals interactions modify atom diffraction such that no closed
form analytical solution has been found.  Hence we developed a
phasor diagram similar to the Cornu spiral to interpret our atom
diffraction data.

In particular, we prove that no combination of grating geometry
and van der Waals interaction strength can cause diffraction
orders to disappear. This is proved using a phasor diagram, and
confirmed experimentally by rotating a diffraction grating about
an axis parallel to the grating bars while measuring the flux in
each diffraction order.

In the standard optical theory of diffraction from a
Ronchi-ruling, i.e. a square-wave absorbing grating, a missing
order is obtained when \begin{equation} n = \pm\frac{ m}{\eta}
\end{equation} where $n$ is the diffraction order, $m$ is an
integer greater than zero, and $\eta$ is the open fraction defined
as window size divided by grating period.  For example, a $50\%$
open fraction suppresses all the even orders.


\begin{figure}
\scalebox{.7}{\includegraphics{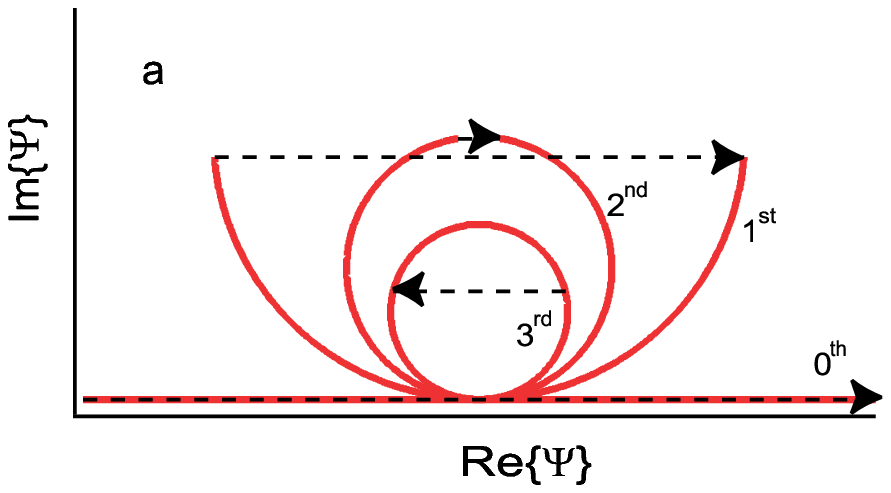}}
\scalebox{.7}{\includegraphics{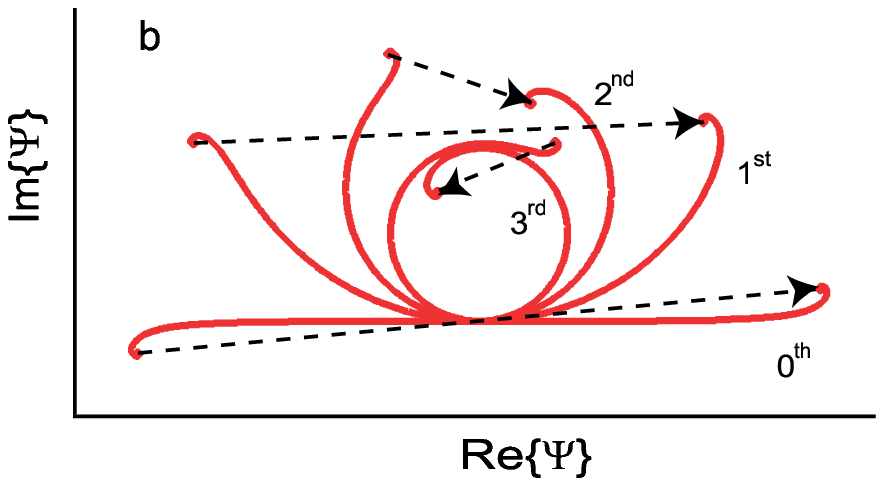}} \caption{\label{fig1} (a)
Phasor diagrams for diffraction into orders $n=0,1,2,3$ from a
grating with open fraction \mbox{$\eta = 0.48$}.  Resultant
vectors are drawn with dashed lines from tip to tail of each
curve.  (b) Van der Waals interactions modify the phasor diagram,
and considerably increase the magnitude of the second order.}
\end{figure}

The origin of missing orders in the optical theory can be
explained with a phasor plot often referred to as a vibration
curve \cite{hr} and shown in Figure \ref{fig1}a. In this approach
a complex amplitude-phase diagram is used to visualize the
amplitude and phase of the field in each diffraction order.
Specifically, we consider the field $\Psi_n$ associated with the
$n^{th}$ diffraction order in the Fraunhauffer approximation due
to a periodic array of apertures illuminated by a plane wave
$\Psi_{inc}$. The slit width $w$ and grating period $d$ determine
the diffraction envelope so the amplitude and phase of each order
is given by:
\begin{equation}\Psi_n=\frac{\Psi_{inc}}{d}\int_{-w/2}^{w/2}d\xi
e^{i\phi_n(\xi)}\label{eqpsi}\end{equation} where $\Psi_{inc}$ is
the incident wave amplitude, $\xi$ is the spatial coordinate
inside an aperture, and $\phi_n(\xi) = n\kappa\xi$ with $\kappa =
\frac{2\pi}{d}$ being the grating wavenumber.

This integral can be visualized by adding phasors of length $d\xi$
and phase $\phi_n(\xi)$ in the complex plane. Curves generated
this way are the real vs imaginary parts of the cumulative
integral of $\Psi_n$ given by Equation \ref{eqpsi}. The magnitude
and phase of a resultant vector (from start to end of such a
curve) correspond to the complex amplitude $\Psi_n$ as shown in
Figure \ref{fig1}. A resultant of zero magnitude represents a
missing order.

Before including phase shifts due to atom-surface interactions,
this integral can be computed analytically:
\begin{equation} \Psi_n =
\Psi_{inc}\textrm{ } \eta \textrm{ sinc}( n \eta) \end{equation}
For comparison with experiment, intensity is given by
$I_n=|\Psi_n|^2$.

In the WKB approximation, to leading order in $V(\xi)$, van der
Waals interactions cause a phase shift $\phi_{vdW}(\xi)$ given by
\begin{equation} \phi_{vdW}(\xi) = \frac{-V(\xi) l }{v \hbar} \end{equation}
where $V(\xi)$ is the van der Waals potential for atoms
interacting with a surface, $l$ is the distance the atom
propagates in the potential, $v$ is the atomic velocity, and
$\hbar$ is Planck's constant divided by $2\pi$. Near a surface the
potential is known to be
\begin{equation}V(r) = \frac{-C_3}{(r)^3}\label{v(r)}\end{equation} where $r$ is the
atom-surface distance \cite{milo94}.  For simplicity, as in
\cite{gris99,gris00,bruh02}  $V(r)$ is approximated by equation
\ref{v(r)} inside a grating window with finite thickness $l$, and
approximated by zero outside the grating window.

\begin{figure}[h]
\scalebox{.75}{\includegraphics{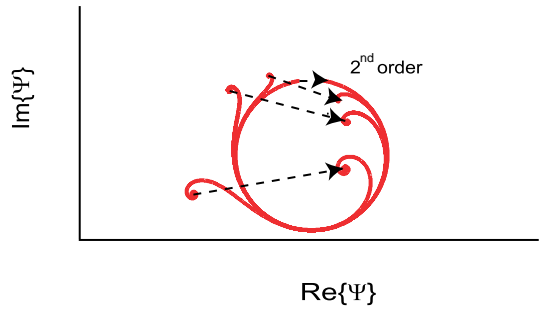}} \caption{\label{fig2}
Phasor diagrams for diffraction from a grating with open fraction
$\eta = 0.48$ shown for diffraction order $n=2$ given different
values of the van der Waals coefficient \mbox{$C_3=0,1,10,100$
eV$\AA^3$.} The resultant vectors from tip to tail are drawn with
dashed lines.}
\end{figure}

Van der Waals interactions with the bars on either side of the
slot thus modify the phase of each point on the phasor diagram.
This phase is described by: \begin{equation} \tilde{\phi}_n(\xi) =
\phi_{n}(\xi) + \phi_{vdW}(\xi) + \phi_{offset}\end{equation}
where $\phi_{n}(\xi)$ is due to the diffraction angle,
$\phi_{vdW}(\xi)$ is due to van der Waals interactions with
surfaces located at $\xi = \pm\frac{w}{2}$, and $\phi_{offset}$ is
a constant added for reasons that will be discussed:
\begin{eqnarray} \phi_{n}(\xi)&= & n\kappa\xi \\
\phi_{vdW}(\xi) &=& \frac{C_3 l}{v \hbar} \textrm{\Large{[}} (\xi
- w/2)^{-3} - (\xi +  w/2)^{-3}\textrm{\Large{]}} \\
\phi_{offset} &=& \frac {16 C_3 l}{v \hbar w^3}. \end{eqnarray}
Examples of phasor diagrams modified by atom-surface interactions
are shown in Figure \ref{fig1}b, and for a range of van der Waals
interaction strengths are shown again in Figure \ref{fig2}.

The constant $\phi_{offset}$ is not physical but is chosen to be
$\phi_{offset} = - \phi_{vdW}(0)$ as a convenience to compare the
shape of the phasor curves for arbitrary van der Waals constant
$C_3$. This phase offset does not change the norm of the resultant
amplitudes, hence it will not affect the proof regarding missing
orders.  It simply rotates the spiral such that
$\tilde{\phi}_n(0)=0$ independent of $C_3$.

For $C_3=0$ the phasor curves are simple circles because the
curvature $\frac{d}{d\xi} \phi_{n}(\xi) = n \kappa$ is constant.
The phase $\phi_{n}(\xi)$ spans an angle $\kappa n w$, and the arc
length of the curve is given by the window size $w$. Thus the
radius of the circle is $\rho = (n\kappa)^{-1}$ and it is centered
at the location $i\rho$. When the curve is an integral number of
full circles, the endpoints overlap and the resultant field has
zero magnitude.  This corresponds to a missing order.

The additional phase due to van der Waals interactions makes the
phasor curve deviate from a circle.  One end of the spiral will
always be $inside$ the circle defined by $\rho$ and the other end
must be $outside$.  This is true because the curvature
$\frac{d}{d\xi} (\phi_{n} + \phi_{vdW})$ increases monotonically
as $\xi$ goes from 0 to $w$/2 and decreases monotonically as $\xi$
goes from 0 to -$w$/2 (and the curvature equals $n\kappa$ at
\mbox{$\xi=0$}). Hence, the two ends of the spiral will never
coincide and the resultant field will never have zero magnitude.

We have now proved regardless of the physical open fraction, there
are never missing orders in atom diffraction from a material
structure unless $C_3 = 0$.  In addition, the diffraction envelope
is no longer described by a $sinc^2$ function as would be the case
if the diffraction were described by a real-valued effective open
fraction.

Our phasor diagram analysis also provides a method to bound the
error on numerically computed amplitudes.  If the limits of
integration in Equation \ref{eqpsi} are replaced by $\pm(w/2 -
\epsilon)$, the maximum error in resultant amplitude is given by
the radius of a circle with the curvature of the phasor diagram at
$\xi = (w/2 - \epsilon)$, i.e. error in $\Psi_n$ is less than
$(\Psi_{inc}/d)R$ where $R^{-1} \equiv d\tilde{\phi}_n / d \xi
\mid _{\xi = (w/2 - \epsilon)}$.

\begin{figure}[h]
\scalebox{.25}{\includegraphics{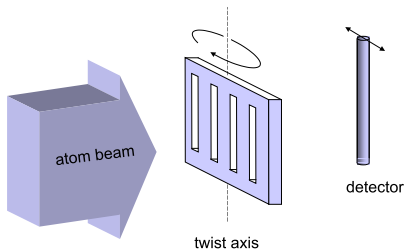}} \caption{\label{fig3}
Experimental geometry.} \end{figure}

\begin{figure}[h]
\scalebox{.6}{\includegraphics*{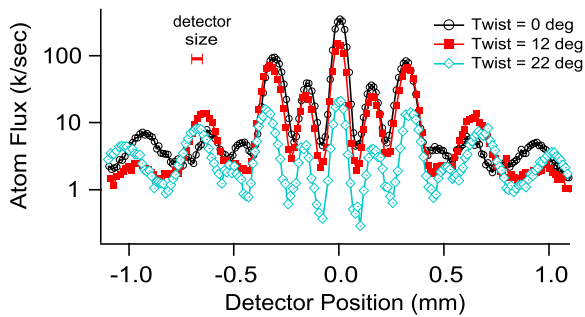}} \caption{\label{fig4}
Diffraction scans with different grating twist.  Diffraction of
both sodium atoms and sodium dimers is visible.  First order atom
diffraction, $I_1$ is located at $\pm 0.3$ mm from the 0$^{th}$
order.} \end{figure}

To confirm this theoretical description, we present atom
diffraction data using a sodium atom beam and a silicon nitride
grating with a period of 100 nm.  We call rotation about a grating
bar $twist$ as shown in Figure \ref{fig3}. Normal incidence
defines zero twist. Diffraction patterns shown in Figure
\ref{fig4} were obtained with the grating held at a twist of 0,
12, and then 22 degrees, by scanning the position of the hot wire
detector. Twist foreshortens the grating period and therefore
slightly increases the diffraction angle. However, the $+n$ and
$-n$ diffraction orders are nearly equally deflected because the
atomic de Broglie wavelength is small compared to the grating
spatial period.

The feature explored in this study is the relative intensity in
each order, which changes considerably as the grating is twisted.
Atom flux diffracted into the zeroth order decreases when the
grating is twisted, but flux into the second and third orders
first increases. Figure \ref{fig4} shows that the second order
intensity is largest for an intermediate twist. Variation in the
relative intensity among diffraction orders is expected because
the projection of the grating viewed from the incident atom beam
changes with twist. However, a model based on absorptive atom
optics is not sufficient to explain our data. Phase shifts due to
van der Waals interactions must be included, as discussed earlier.

\begin{figure}[]
\scalebox{.6}{\includegraphics*{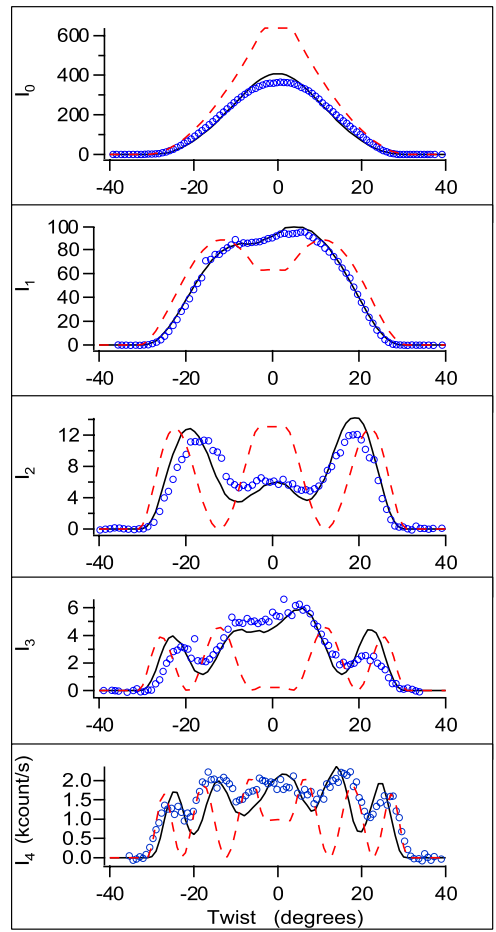}} \caption{\label{fig5} Data
(circles) and models (dashed and solid lines) of the intensity in
each diffraction order as a function of grating twist. The model
parameters are: $d$=100nm, w=67nm, $l$=116nm, and
$\alpha=3.5^{\circ}$.  For dashed lines $C_3$=0 (using equation
\ref{weights}), and for solid lines $C_3$=5 eV$\AA{^3}$ (using
equation \ref{Psi_vdW}). Statistical error bars for each data
point are smaller than the circles.} \end{figure}

We also present data obtained while scanning the grating twist and
leaving the detector at one position (Figure \ref{fig5}). In the
latter experiment we have measured $I_n=|\Psi_n|^2$ for each
n=[0,4] while continuously changing the projected open fraction.
This technique is well suited to the task of searching for missing
orders, because the projected open fraction can be scanned through
$\frac{1}{2}$, $\frac{1}{3}$, $\frac{1}{4}$, i.e. values that
would make missing orders according to the optical theory.

\begin{figure}[h]
\scalebox{.3}{\includegraphics{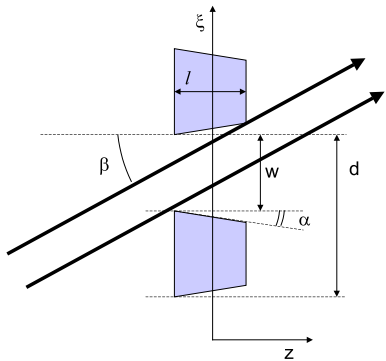}} \caption{\label{fig6} Top
view of the atom beam passing through the grating slots.  The
grating twist is denoted by $\beta$, the wedge angle of each bar
is $\alpha$, the grating period is $d$, the thickness is $l$, and
physical open width is $w$.}
\end{figure}

For a perfectly thin grating, twist would not affect the open
fraction.  For our gratings, with a geometry shown in Figure
\ref{fig6}, twist does in fact modify the projected open fraction.
Furthermore, due to trapezoidal grating bars, van der Waals phase
shifts must be carefully analyzed.  With reference to Figure
\ref{fig6}, twisting the grating by angle $\beta$ causes slots to
appear narrower so the resulting open fraction is:
\begin{equation}
\begin{split}
\eta(\beta) &=
    \begin{cases}
 \frac{w}{d} &; |\beta| < \alpha  \\
 \frac{w-l(\tan\beta - \tan\alpha)}{d} &; \alpha < |\beta| < \beta_{max} \\
 0 &;  \beta_{max} < |\beta|
    \end{cases}
\end{split}
\label{of}\end{equation}

where $w$ is the slot width viewed at normal incidence, \mbox{$l$
is} the thickness of the grating, $\alpha$ is the wedge angle of
the bars, $\beta$ is the twist, $\beta_{max}$ is the maximum twist
at which any flux is transmitted, and $d$ is the grating period
viewed at normal incidence. The intensity in different diffraction
orders then depends on twist as: \begin{equation} I_n(\beta) =
\eta(\beta)^2 \textrm{sinc}^2\textrm{\Large{(}}\eta(\beta)
n\textrm{\Large{)}} \label{weights} \end{equation} This model
without an allowance for atom-surface interactions was used to
predict intensity in the first five orders as a function of twist
shown in Figure \ref{fig5} (dashed lines), and compares poorly
with the data.

Two features of this model are familiar from standard diffraction
theory. First, in the limit of small open fraction the intensity
of each order becomes equal. This happens at large twist.  Second,
the orders are completely suppressed at certain angles for which
$n=m\eta(\beta)^{-1}$, i.e. missing orders are predicted. However,
the atom beam flux is never entirely suppressed until the
projected open fraction is zero, i.e. there are no missing orders
in the data.

Several of the grating geometry parameters are known from the
manufacturing process and SEM images \cite{sava96}. The period is
$d$ = 100 nm and the window size is $w$ = 67$\pm5$ nm, so at
normal incidence the open fraction is approximately \mbox{67\%}.
The only physical grating parameters left in the optical model are
the thickness, $l$, and wedge angle, $\alpha$. These are
constrained by measuring the maximum twist at which any flux is
transmitted, and must satisfy the condition $w=l(\tan\beta_{max}-
\tan\alpha)$. The data show $\beta_{max} = 31 \pm 2^{o}$.  With
this constraint, there is only one free parameter in the optical
model, i.e. $\alpha$.

The relative intensity for each order is determined in the optical
model by Eq \ref{weights}, and can be compared to the data that
were recorded without changing the detector gain or incident beam
flux. However, the velocity distribution of the beam broadens the
higher order diffraction peaks. Each diffraction order has a HWHM
in the detector plane given approximately by $\sigma_n^2 =
\sigma_0^2 + (\Theta_n L \frac{\sigma_v}{v})^2$ where $\sigma_0$
is the HWHM of the zeroth order, $L$ is the distance between the
grating and the detector, $\frac{\sigma_v}{v}$ is the ratio of the
HWHM spread in velocity to the average velocity, and $\Theta_n$ is
the diffraction angle \cite{gris00}. The velocity ratio is 1/15
and the average velocity is 1000 m/s. To allow for the velocity
distribution, the relative intensity of each diffraction order is
multiplied by $\sigma_n^{-1}$ in the model. We have not accounted
for the change in $\Theta_n$ with twist, that has the effect of
further reducing the recorded intensity of the higher orders at
large twist.

To include the van der Waals interactions with the twisted
grating, the model must take account for the varying distance to
the surface as atoms pass through the slots as shown in Figure
\ref{fig6}, i.e. the transverse coordinate in the grating now
depends on the longitudinal position $\xi \rightarrow \xi_0 +
\xi(z)$. The potential due to each interior wall is approximated
by the van der Waals potential for an infinite surface when the
atoms are inside the grating slots, and zero elsewhere. Then the
phase shift due to one wall of a twisted grating is:
\begin{equation}  \phi_{vdW}(\xi) = \frac{-2C_3 \textrm{\huge{[}}
(x\tan(\theta)-w/2+\xi_0)^{-2} \textrm{\huge{]}}
\textrm{\huge{$\mid$}}^{x=l/2}_{x=-l/2}}{v \hbar\tan(\theta)
\cos(\beta)} \label{phi_vdW} \end{equation} where
$\theta=\beta\pm\alpha$ with the sign depending on whether the
classical paths get closer or farther from the grating wall as a
function of $z$.  Changing the sign of $w$ describes the phase
shift from the other wall.  As before, the field amplitude in
$n^{th}$ order diffraction is given by an integral, but now the
limits of integration depend on grating twist as shown in Equation
\ref{of} as does $\phi_{vdW}$ given by Equation \ref{phi_vdW}:
\begin{equation} \Psi_n =
\int_{\eta(\beta)\frac{d}{2}}^{\eta(\beta)\frac{d}{2}} e^{i(
\phi_{n}(\xi)+\phi_{vdW}(\xi))} d \xi. \label{Psi_vdW}
\end{equation}

Equation \ref{Psi_vdW} can now be used to describe intensity in
each order as a function of van der Waals coefficient, atom
velocity, and grating twist: $I_n(C_3,v,\beta)= |\Psi_n|^2$. When
$C_3$ is zero, the expression for the intensities reduces to
Equation \ref{weights}. In comparison, when $C_3$ is not zero the
model predicts no missing orders and agrees qualitatively with the
data in Figure \ref{fig5} (this model is shown with solid lines).
We note the van der Waals interaction diverts flux from the zeroth
order into the higher orders, and tends to smooth the features
given by Equation \ref{weights}. Even the slight asymmetry in
intensity as a function of twist is reproduced.  To our knowledge
this is the first hint of a fabricated structure acting as a
blazed grating for atom waves.

When used to measure the strength of $C_3$ for sodium atoms and a
grating made of silicon nitride, we determine a value for $C_3 = 5
^{+5}_{-2} $eV$\AA^3$.  Further work on the precise shape of the
grating, and the van der Waals potential in all space due to the
structure is needed to reduce this uncertainty.

The maximum asymmetry we observe in the first order occurs at a
twist of $\pm 5 ^o$ and is $I_1^{+5^o}/I_1^{-5^o} = 1.1$.  In
simulations with a larger wedge angle and larger $C_3$, the
asymmetry can be as large as 1.5.  An asymmetric distribution of
intensity between the $+1^{st}$ and $-1^{st}$ orders, as this
implies, would be useful for atom interferometers that only employ
the $0^{th}$ and $+1^{st}$ diffraction orders.

Even at normal incidence the van der Waals interaction has reduced
the intensity of the zeroth order diffraction by a factor of 0.65
compared to the zeroth order flux predicted with $C_3=0$. As an
extrapolation, if we could obtain similar gratings with a 20 nm
period, the flux transmitted into the zeroth order will be reduced
to 0.06 of the flux predicted with $C_3=0$. Hence, van der Waals
interactions pose a significant problem for sub-100-nanometer
scale atom optics.

In conclusion, a novel way to measure the atom-surface interaction
potential was presented. By twisting a 100 nm period diffraction
grating, we show that atom-surface interactions prevent missing
orders, and cause asymmetric diffraction patterns. Both
observations are explained by a complex transmission function and
a phasor analysis similar to the Cornu spiral.

We are indebted to T. Savas and H.I. Smith for fabrication of the
100 nanometer period material gratings \cite{sava96}.  We also
thank B. Anderson for a critical reading of this manuscript and
both H. Uys and P. Hoerner for technical assistance.

\end{document}